# Optical activation and detection of charge transport between individual color centers in room-temperature diamond

Artur Lozovoi[1], Harishankar Jayakumar[1], Damon Daw[1], Gyorgy Vizkelethy[3], Edward Bielejec[3], Marcus W. Doherty[4], Johannes Flick[5], Carlos A. Meriles[1,2,*]

**Charge control of color centers in semiconductors promises opportunities for novel forms of sensing and quantum information processing. Here, we articulate confocal fluorescence microscopy and magnetic resonance protocols to induce and probe charge transport between discrete sets of engineered nitrogen-vacancy (NV) centers in diamond, down to the level of individual defects. In our experiments, a 'source' NV undergoes optically-driven cycles of ionization and recombination to produce a stream of photo-generated carriers, one of which we subsequently capture via a 'target' NV several micrometers away. We use a spin-to-charge conversion scheme to encode the spin state of the source color center into the charge state of the target, in the process allowing us to set an upper bound to carrier injection from other background defects. We attribute our observations to the action of unscreened Coulomb potentials producing giant carrier capture cross-sections, orders of magnitude greater than those typically attained in ensemble measurements. Besides their fundamental interest, these results open intriguing prospects in the use of free carriers as a quantum bus to mediate effective interactions between paramagnetic defects in a solid-state chip.**

Long seen as a hurdle to semiconductor technologies, point defects in solids have presently emerged as a promising platform for nanoscale sensing and quantum information processing in the solid state[1,2]. Although the list of defects with attractive physical properties is growing rapidly[3], select spin-active color centers in wide-bandgap semiconductors such as diamond[4], silicon carbide[5], or hexagonal boron nitride[6] are presently garnering broad attention, mostly due to their favorable optical and/or magneto-optical properties. Given the profound impact of the charge state on defect's spin and photon emission properties, a broad effort is being devoted to more fully understand defect photo-ionization[7] as well as carrier trapping[8], both deep inside the crystal[9-11] and near the surface[12,13]. Further, charge control methods are being explored as a route towards reduced spectral diffusion[14,15], alternative forms of electric field sensing[16], super-resolution imaging[17,18], or for Stark tuning of the optical transitions through photo excitation of trapped charges[19]. Electroluminescence produced by electron-hole recombination at individual color centers has been observed at room temperature in an all-diamond diode structure[20].

Of special interest is the interplay between the color center's spin and charge states, already exploited, e.g., to enhance detection sensitivity[21-23], create decoherence-protected qubit states[24], and store and retrieve classical information[25]. Recent experiments made use of spin-to-charge conversion protocols to demonstrate photo-electrical detection of magnetic resonance from individual color centers in diamond[26-28] and silicon carbide[29]. On a related front, spin-based electrometry from a probe color center was used to monitor changes in the local charge environment[30]. Spin-encoded photo-generated carriers propagating from source to target point defects have also been proposed as a bus to communicate pairs of distant qubits in a solid-state chip[31].

Here, we make use of high-energy ion implantation to engineer spatial patterns of nitrogen-vacancy (NV) centers in pristine bulk diamond. With the help of multi-color confocal microscopy and optically detected magnetic resonance, we investigate the photo-ionization and carrier trapping properties of small NV sets, down to the limit of individual color centers. We focus on the transport of carriers between physically separate 'source' and 'target' defects, which we address selectively via independent laser beams. Further, we use spin-dependent ionization to filter out carriers stemming from source NVs (as opposed to other concomitant defects in the sample) and show they are dominant in establishing carrier transport. Bringing down the number of source and target defects to individual NVs, we leverage our ability to probe inter-defect carrier transport as a means to directly measure NV carrier trapping cross-sections.

## Results

**Carrier transport between individual color centers.** We conduct the present experiments in a Type 2a diamond with an intrinsic nitrogen concentration of ≲ 5 ppb. To engineer NV centers at select locations, we first use a focused ion beam from a tandem accelerator to implant nitrogen ions ~10 μm deep into the crystal (Fig. 1a), which we subsequently convert into NV centers via sample annealing[32,33]. Carriers photo-

---

[1]Department. of Physics, CUNY-City College of New York, New York, NY 10031, USA. [2]CUNY-Graduate Center, New York, NY 10016, USA. [3]Sandia National Laboratories, Albuquerque, New Mexico 87185, USA. [4]Laser Physics Centre, Research School of Physics, Australian National University, Canberra, Australian Capital Territory 0261, Australia. [5]Center for Computational Quantum Physics, Flatiron Institute, New York, NY 10010, USA. *E-mail: cmeriles@ccny.cuny.edu

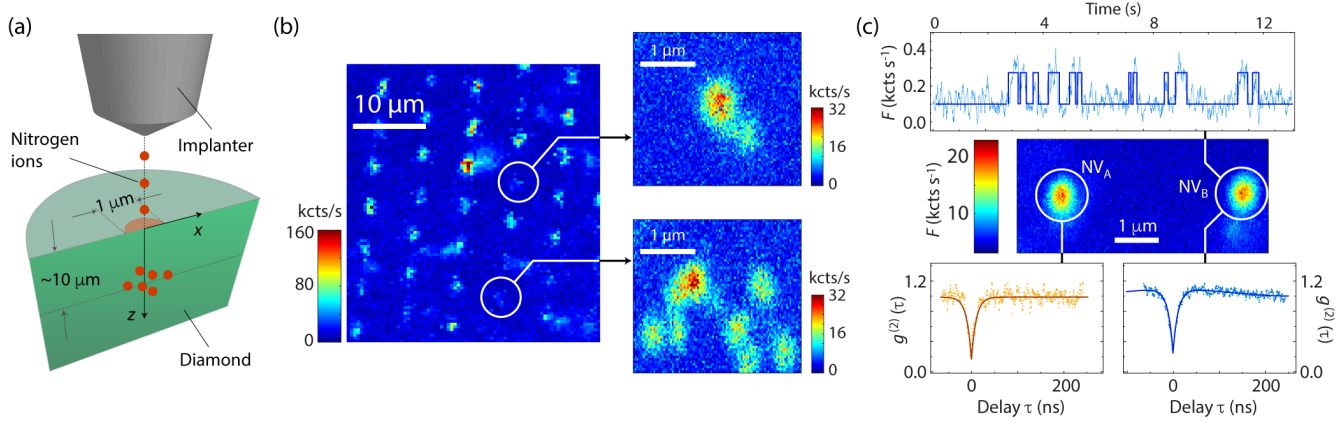

**Fig. 1 | Engineering NV spatial patterns.** (a) Ion implantation geometry. We accelerate $N^+$ ions using a tandem accelerator focused over a ~1 μm diameter spot on the diamond surface. (b) NV fluorescence image of implanted grid. Zoomed images on the right show individual NVs. (c) Confocal fluorescence image of individual NV centers 4 μm apart. In our experiments, photo-generated carriers diffuse from $NV_A$ to $NV_B$ (respectively referred to as the 'source' and 'target' NVs). The lower inserts display the individual photon autocorrelation curves $g^{(2)}(\tau)$. The respective two-color fluorescence blinking from NVs, expected for single photon emitters, injected from defects at this depth remain within the bulk crystal during inter-defect propagation, thus making our observations free from surface effects. Further, the ability to focus the ion beam over ~1-μm-diameter spots gives us the opportunity to engineer virtually arbitrary implantation geometries (see Methods and Extended Data Fig. 1). Fig. 1b shows a simple case, here in the form of an NV grid; by varying the ion beam fluence, we produce arrays with a variable color center content often leading to the formation of pairs of NVs micrometers apart. Fig. 1c shows one such example, as demonstrated by photon autocorrelation and fluorescence time trace measurements at each site.

An initial demonstration of carrier propagation between individual color centers is presented in Fig. 2. In these experiments carrier photo-generation derives from NV charge state interconversion during prolonged green 520 nm optical excitation (Fig. 2a); electron injection into the conduction band emerges from $NV^-$ ionization whereas holes are produced during $NV^0$ recombination[8,9]. Using suitable bandpass filters to separate $NV^-$ from $NV^0$ emission[34], we expose inter-defect carrier propagation — here relying on self-diffusion, see below — by monitoring the fluorescence from NV$_B$ (Fig. 2b): Initially prepared in the negatively charged, 'bright' state, the target NV turns to 'dark' upon capture of a hole produced by NV$_A$ (here acting as the carrier source, Fig. 2c). Note that although the same (or comparable) number of electrons and holes is produced during a green laser park, Coulomb attraction makes the hole capture cross section of NV$^-$ dominant[8], thus leading to a net fluorescence change. Further, given the relatively low defect content, this change is infinitely long-lived in the dark[25], and thus can be subsequently imaged with the help of a 594 nm beam (orange illumination excites only NV$^-$ fluorescence[7], see diagram in Fig. 2a). Note that since some NV$^0$ recombination takes place at 594 nm (see Methods), the probe intensity is chosen so as to optimize the photon count without impacting the NV charge state.

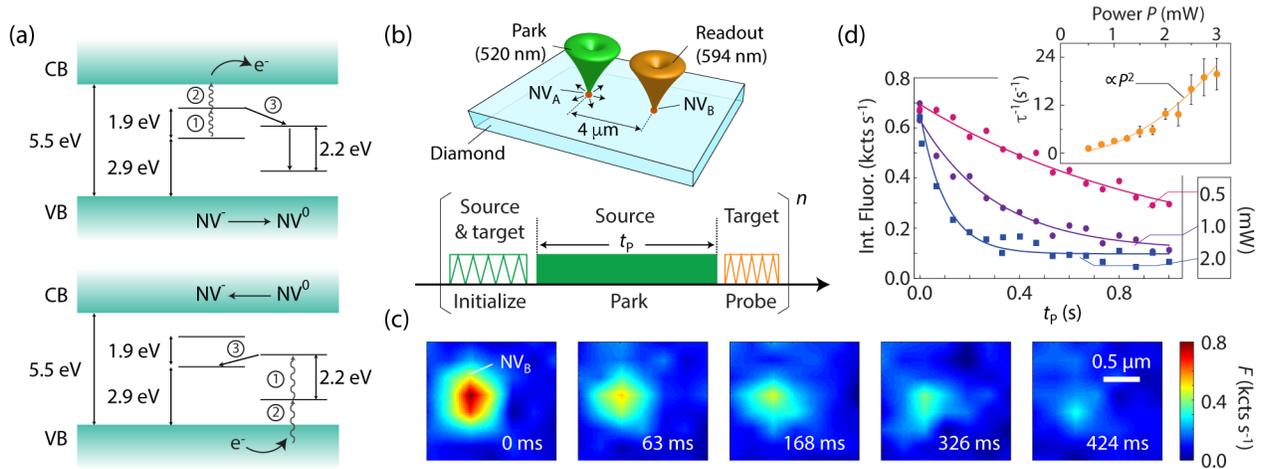

**Fig. 2 | Controlling charge transport between source and target color centers.** (a) Schematics of NV photochromism under optical (~460−637 nm) excitation. The upper (lower) panel depicts $NV^-$ ionization ($NV^0$ recombination). In each case, steps (1) through (3) describe the successive absorption of two photons followed by NV charge transformation and injection of an electron (hole) into the conduction (valence) band (CV and VB, respectively). (b) Experimental geometry (top) and charge transfer protocol (bottom); the green (orange) zig-zag indicates a charge initialization (readout) scan at 520 nm and 2.5 mW (594 nm and 7 μW). The solid green block indicates 520 nm, 2.5 mW laser parking at $NV_A$. (c) Fluorescence image of $NV_B$ upon application of the protocol in (b) for variable park durations $t_P$ (lower right corner). Each image results from a 7×7-pixel, 594 nm beam scan (7 μW, 10 ms per pixel); for clarity, we use linear color interpolation between sites. (d) Integrated fluorescence in a 1.5×1.5 μm$^2$ area containing $NV_B$ as a function of $t_P$ for variable park powers; solid lines indicate exponential fits. (Insert) Unit time charge conversion probability $\tau^{-1}$ of $NV_B$ as a function of the park power at $NV_A$. Error bars indicate standard deviation. In these experiments, the total number of repeats is $n = 10^2$.

Fig. 2d shows the integrated fluorescence in the area surrounding NV$_B$ as a function of the park time $t_P$ for some example green laser powers. For sufficiently long park times, the integrated fluorescence at the target site decays to background levels, consistent with prior observations in ensembles[9]. Capitalizing on the photoluminescence decay curves to extract the NV$^-$ hole capture rates $\tau^{-1}$ at various powers, we find a quadratic dependence (insert in Fig. 2d), as expected for the two-photon processes governing charge interconversion in the source NV[7] (Fig. 2a, see also Extended Data Fig. 2). In each case, the time scale $\tau \approx 4\pi d^2/(k_h \sigma_h)$ of the process — in the order of a few hundred milliseconds depending on laser power — arises from an interplay between the target hole capture cross section $\sigma_h$ (see below), the inter-defect distance $d$ (~4 µm in this case), and the unit time number of photo-generated holes $k_h$ (produced by the source NV as it cycles from negative to neutral and back under continuous green illumination). The latter can be related to the NV ionization and recombination rates as $k_h = (k_{ion}^{-1} + k_{rec}^{-1})^{-1}$. Since $\sigma_h/4\pi d^2$ is the fractional area associated with the target NV, $\tau$ can be intuitively interpreted as the time during which enough holes have been generated to ensure a capture event with near 100% probability.

We note that the time-averaged fluorescence from NV$_A$ remains unchanged as a function of the illumination duration or laser power, consistent with stochastic cycles of ionization and recombination under continuous green excitation (see Extended Data Fig. 3a-3b). By contrast, red excitation of the source induces rapid conversion of NV$_A$ into the neutral, dark state with no bleaching in the fluorescence from NV$_B$, anticipated for a one-directional ionization process that entails no hole injection (Extended Data Fig. 3c-3d). Combined, these observations suggest that photo-generated carriers originate predominantly from NV$_A$ (as opposed to other neighboring defects), an important consideration we address more in detail immediately below.

**Spin-to-charge conversion as a carrier source filter.** Although a large fraction of implanted nitrogen ions converts into NV centers after sample annealing[33], residual defects — e.g., in the form of substitutional nitrogen or 'P1 centers' — may coexist, arguably serving as parasitic charge injection sources. To selectively filter out contributions stemming from NV centers alone, we implement the protocol in Fig. 3a:

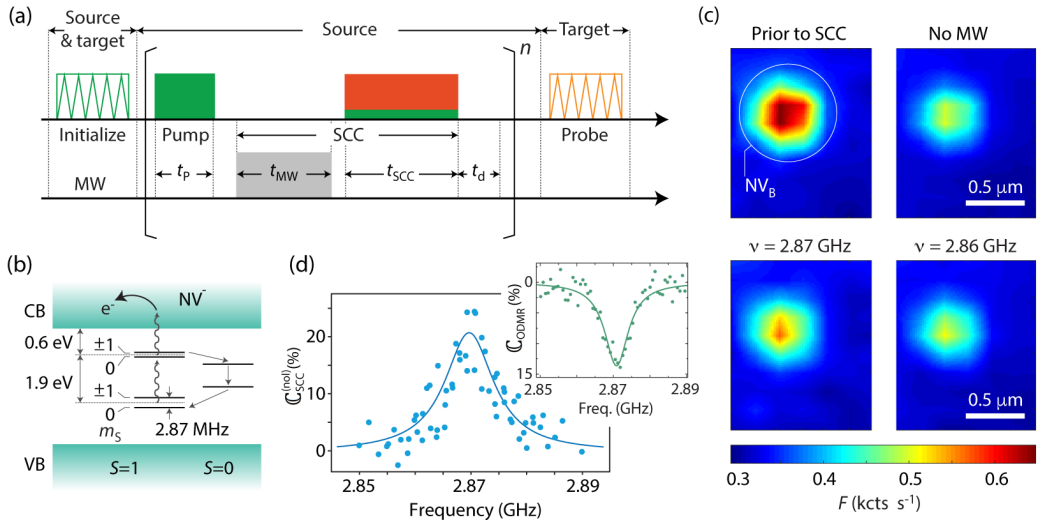

**Fig. 3 | Filtering out NV-generated carriers.** (a) Spin-to-charge conversion (SCC) protocol. Green/red solid blocks indicate local laser excitation at 520 nm and 632 nm of NV$_A$ (the source NV in this case); green and orange zig-zags are laser scans at 520 nm and 594 nm, respectively. (b) NV$^-$ energy diagram including the zero field splittings of the ground and excited state triplets ($S = 1$) as well as the singlet ($S = 0$) manifold levels. Intersystem crossing is more likely from the $m_S = \pm 1$ states in the excited triplet, thus leading to a spin dependent NV$^-$ ionization rate. (c) (Top) Confocal images of NV$_B$ (the target NV) before and after application of the protocol in (a) (left and right panels, respectively) in the absence of MW. (Bottom) On-resonance excitation at 2.87 GHz (left panel) reduces carrier generation at the source NV and thus diminishes bleaching of the target. For clarity, we use linear color interpolation between sites. (d) Integrated fluorescence contrast ℂ at the target NV as a function of the MW frequency. The insert in the upper right shows the optically-detected magnetic resonance spectrum of the source NV. In these experiments, $t_P = 1$ µs, $t_{SCC} = 0.3$ µs, $t_{MW} = 0.15$ µs, $t_d = 10$ ns, and $n = 5 \times 10^5$; the red laser power during SCC is 16 mW while the green laser power is 200 µW during the initialization scan, $t_{SCC}$, and $t_P$. All other conditions as in Fig. 2.

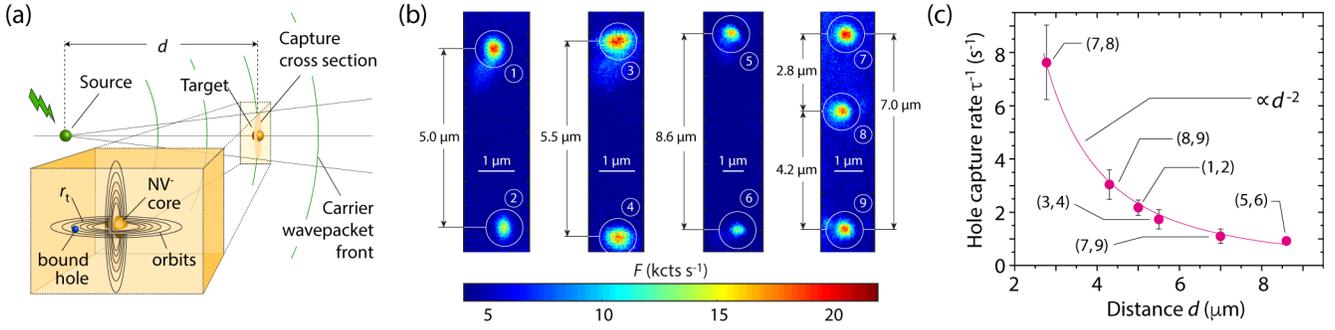

**Fig. 4 | Photo-induced carrier transport between NVs at variable distances.** (a) Schematics of carrier scattering dynamics. Assuming an initial wave packet centered at the source, the carrier capture probability decreases approximately with the square of the distance $d$ to the target. Hole capture takes place via the formation of a transient bound exciton characterized by the trapping radius $r_t$. (b) To gauge the impact of the distance on the carrier capture dynamics, we select a pair of target and source NVs from a discrete set, as shown in the fluorescence images (white circles). (c) Hole capture rate $\tau^{-1}$ as a function of the distance $d$; the solid line indicates an inverse square dependence. Indices in each pair denote the source and target NV sites in each measurement following the labels in (b); the green laser power during these experiments is 1 mW, all other conditions as in Fig. 2.

Following charge and spin initialization, we subject the source NV to cycles of microwave (MW) excitation and simultaneous green (520 nm) and red (632 nm) excitation with relative powers adapted for optimal NV⁻ ionization[35,38]. Similar 'spin-to-charge' conversion (SCC) schemes have been utilized successfully in the recent past, both in individual NVs and small ensembles[22,35,36]. Briefly, resonant MW driving of the zero-field transition in the NV ground electronic state (where the spin number is $S = 1$) counters optical spin pumping into the $m_S = 0$ projection of the triplet, hence increasing the chance of intersystem crossing into the singlet manifold (Fig. 3b). The associated shelving that accompanies this crossing impacts the timing in the optical excitation cycle of NV⁻ and thus effectively lowers its two-step ionization rate[26-29]. Correspondingly, there is a reduction in the number of photo-generated holes approaching the target color center and thus a relative increase in its emitted fluorescence.

We isolate the impact of the NV$_A$-produced carriers by referencing the NV$_B$ fluorescence to the response with no MW. To quantify the effect, we measure the non-local SCC contrast $\mathbb{C}_{SCC}^{(nol)} = \Delta F(\nu)/\Delta F_{Ref}$, where $\Delta F(\nu) = F_{On}(\nu) - F_{Off}$ is the fluorescence change with and without MW at frequency $\nu$, and $\Delta F_{Ref} = F_{Ref} - F_{Off}$ with $F_{Ref}$ denoting the target NV fluorescence in the absence of optical excitation at the source. Experimentally, we find maximum contrast when the MW is on resonance with the NV⁻ crystal field transition (Fig. 3c); by monitoring the response as a function of the MW frequency, we reconstruct what amounts to an ancilla-detected NV⁻ spin resonance spectrum[37] (Fig. 3d). Unlike in regular optically detected magnetic resonance experiments (insert in Fig. 3d), the fluorescence change is positive, consistent with the effectively lower carrier photo-generation at the source. Since the local SCC contrast in NV$_A$ reaches up to 30% (see Extended Data Fig. 4), the observed fluorescence change in NV$_B$ indicates that not less than 75% of the trapped photo-generated carriers stem from the source NV. We emphasize this is a lower bound because, e.g., background holes from NV$_A$ produced during the spin initialization time $t_P$ can lower the observed spin contrast. For completeness, we mention that virtually identical results are derived from carrier transport experiments between islands each comprising few NVs (Extended Data Fig. 5). Externally applied fields can be exploited to gain some control on the transport process (see Methods and Extended Data Fig. 6) although the formation of space charge fields can substantially complicate the dynamics at play[38].

**Modeling the capture process.** An interesting pending question relates to the impact of the inter-defect distance $d$ on the carrier capture probability. In the simplest picture, the probability distribution describing photo-generated carriers evolves radially from the source NV (Fig. 4a), thus making the carrier capture probability proportional to $d^{-2}$. To test this notion, we use the protocol in Fig. 2b to monitor carrier transport between pairs of NVs separated by variable distances ranging from 2.5 to 9.5 μm (Fig. 4b). Fig. 4c shows the results for a fixed park laser power and duration confirming the anticipated inverse square dependence. From the expression $\tau \approx 4\pi d^2/(k_h \sigma_h)$ introduced above and using $k_h = (k_{ion}^{-1} + k_{rec}^{-1})^{-1} \approx 2 \times 10^5$ s⁻¹ at the applied green laser power (1 mW), we find $\sigma_h \approx 3 \times 10^{-3}$ μm²; for the NV pair in Fig. 2, this roughly amounts to the equivalent of 1 capture out of $10^5$ generated holes (see Methods).

To rationalize our observations, we describe hole capture as a process that involves the formation of a bound exciton state featuring an electron localized at the NV core and an orbiting hole[20]. In this model — recently invoked to interpret the absorption spectrum of neutral silicon-vacancy centers in diamond[39] — eigenstates follow a hydrogenic sequence whose radii $r_n$ grow with the square of the orbital quantum number $n$. Ab initio calculations support this model, although size limitations in the supercell restrict our analysis to the lower energy orbital (whose energy slightly departs from that predicted by the hydrogenic series due finite size of the charge distribution at the core, see Methods and Extended Data Figs. 7 and 8).

Physically, hole capture takes place via a cascade process involving the consecutive emission of phonons[40]. Asking that the exciton's binding energy be comparable to the free hole thermal energy $\kappa_B T$ — ~25 meV in our room-temperature experiments — we derive the trapping radius $r_t = e^2/(4\pi\varepsilon\kappa_B T) \approx 10$ nm, where $e$ denotes the elementary charge and $\varepsilon = 5.7\varepsilon_0$ is the dielectric permittivity of diamond; correspondingly, we estimate the Onsager capture cross section as $\sigma_{h,On} \sim \pi r_t^2 \approx 3\times10^{-4}$ μm$^2$.

One possible key to the giant cross section observed herein — several orders of magnitude greater than typical values considered in the literature[41] — is the lack of thermally activated carriers, which leads to the unscreened action of the Coulomb potential from a point defect. Equally important is the high sample purity, favoring the transient formation of bound exciton orbitals with large radii. In a semi-classical picture, hole capture starts immediately after orbital formation via phonon scattering processes in which the approaching carrier releases part of the kinetic energy gained as it accelerates towards the charged defect. For the present experimental conditions, the Langevin regime — where multiple collisions occur within the trapping radius — seems more appropriate because the carrier mean-free path in room-temperature diamond amounts to ~15 nm[42,43], comparable to $r_t$. Using the capture cross section formula valid in this limit[40], we attain $\sigma_{h,La} = 4.35\times10^{-3}$ μm$^2$, close to the experimental value. We warn this agreement can be partly fortuitous as the Langevin formula amounts to a crude approximation.

**Discussion**. Transport of charge carriers between individual color centers as reported here promises novel opportunities, both fundamental and applied. For example, extending these experiments to include spin initialization of the target NV or the photo-generated carrier[31] could be leveraged to address the question of spin selectivity during trapping, or to determine the free carrier spin lifetime (still unknown despite the near ideal properties of diamond as a spintronics material[44,45]). Similarly intriguing are local optical spectroscopy measurement designed to probe the higher energy orbitals of a neutral NV, especially when we note that Rydberg states featuring 2 μm radii (corresponding to $n = 25$) have already been observed in $Cu_2O$[46].

Similarly attractive are extensions to low temperatures (where spin-to-charge conversion can be implemented with near-unit fidelity[47]), or in the presence of externally applied electric fields (provided space charge effects can be used to an advantage[38]). Further, by simultaneously monitoring the target and source fluorescence, it should be possible to time-correlate carrier photo-generation and capture. This class of 'heralding' — applied to NVs or adapted to other neighboring defects such as the P1 center — could find use as a strategy towards distributing entanglement between remote spin qubits[31]. Note that increased hole capture probabilities could be attained by carrier confinement[48] (e.g., via diamond nano-beams) or through the use of carrier lensing[49] (in the limit of ballistic propagation). More generally, the ability to encode the spin state of a source qubit in the charge state of an ancilla defect (not necessarily paramagnetic) could prove useful to enhance detection sensitivity of point defects with low quantum yield (e.g., rare earth ions[37,50]) or with photon emission at impractical wavelengths (e.g., the SiV$^0$ in diamond[51]).

**Methods**

**Experimental setup**

The optical measurements throughout this work use a home-built confocal microscope with an oil-immersion objective (NA=1.3) equipped with three continuous wave (cw) diode lasers with emission at 520 nm, 632 nm, and 594 nm (Coherent); in all cases, we reach a diffraction limited illumination spot (with diameter of order 0.5 μm). All three laser beams are combined with the use of two dichroic mirrors (605 nm long-pass and 550 nm long-pass from Semrock) into a single-mode fiber (Thorlabs). The 520 nm and 632 nm lasers allow for fast pulsing (100 MHz), whereas the 594 nm beam is controlled via an acousto-optic modulator with <30 ns switching times (Isomet). A single-photon counting module based on an avalanche photodiode (Excelitas) is used for collection. A 650 nm long-pass dichroic mirror (Semrock) separates the excitation light from the detected photoluminescence, which is then coupled into a 25 μm multi-mode fiber (Thorlabs). This configuration proved to yield a higher count rate and sensitivity for the charge-state-based measurements presented herein, as compared to a collection path with a single-mode fiber. Spin resonance microwave transitions are addressed through an omega-shaped strip-line antenna imprinted on an electronic board, also serving as the sample support. We displace the objective along the optical axis via a high-precision (<10 nm) piezoelectric nano-positioner (Edmund Optics). We use a time-correlated single-photon counting unit PicoHarp 300 (PicoQuant) in Hanbury Brown-Twiss configuration for autocorrelation measurements. All experiments are carried out under ambient conditions.

**Sample preparation**

We conduct our experiments in an electronic-grade [100] diamond crystal (2×2×0.2 mm$^3$) purchased from DDK with intrinsic N content of $\lesssim$ 5 ppb (which corresponds to an inter-nitrogen separation of $\gtrsim$ 200 nm). To engineer NV centers at select locations we use a tandem ion accelerator at Sandia National Laboratory (New Mexico, USA) producing a 20 MeV $^{14}$N beam with a ~1-μm-diameter focal spot (Extended Data Fig. 1a). Implantation patterns of different sizes, shapes, and fluences were created using this system; some examples are shown in Extended Data Figs. 1b–1d. From Stopping and Range Ions in Matter (SRIM) calculations[52] (Extended Data Fig. 1e), we estimate ions propagate for about 12 μm into the crystal before coming to a stop; the modelled lateral straggle amounts to 0.7 μm (a value that must be convoluted with the beam size, not considered here). The NV depth resulting from high-energy ion implantation ensures that charges propagating between NVs are not affected by surface effects (e.g., scattering or trapping by surface defects).

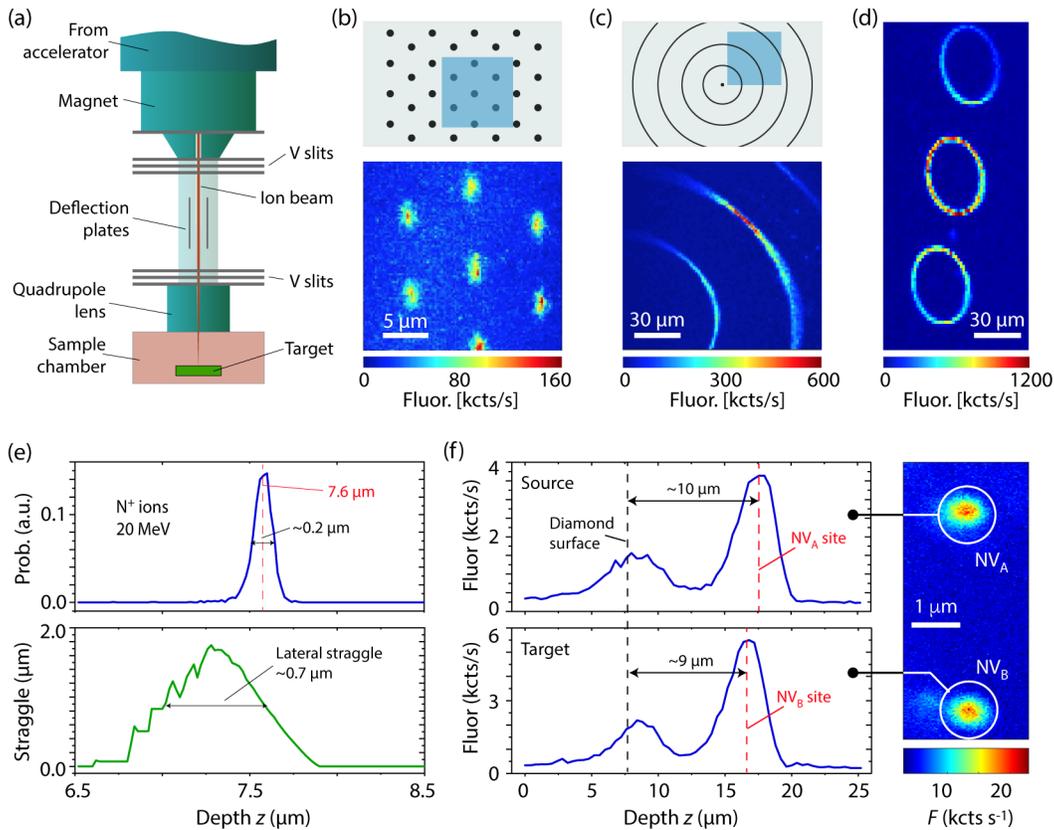

**Extended Data Fig. 1 | Engineering deep NV arrays via a high-energy focused ion beam.** (a) Schematics of the ion implanter. (b-d) Example NV patterns; fluorescence images in (b) and (c) correspond to the shaded areas in the larger NV arrays (upper schematics). (e) SRIM simulations assuming an implantation energy of 20 MeV; the top and bottom plots are the number probability distribution and lateral ion straggle as a function of depth, respectively. (f) NV fluorescence intensity as a function of depth for the source and target single NVs in Fig. 1c of the main text (top and bottom plots, respectively).

We set the beam fluence within the range $1\times10^8$–$5\times10^{11}$ ions/cm$^2$, with the lowest bound corresponding to 2 ions per implantation spot (aimed at producing single-NV sites). Following implantation, we implement a six-step sample annealing protocol[32], namely, (*i*) a 20-400 °C ramp-up during 1 h, (*ii*) a 400 °C anneal for 4 h, (*iii*) a 400-800 °C ramp-up during 1 h, (*iv*) an 800 °C anneal for 2 h, (*v*) an 800-1200 °C ramp-up for 1 h, and (*vi*) a 1200 °C anneal for 2 h. Throughout the process, we maintain the vacuum level in the annealing chamber below $10^{-6}$ Torr. Subsequently, we treat the sample with a tri-acid mixture (nitric, sulfuric, perchloric acids in 1:1:1 proportion) for 1 hour to eliminate graphite and other impurities formed on the surface, as well as to provide oxygen termination.

A direct measurement of the fraction of nitrogen implants converting into NVs proved difficult because space charge fields during implantation created instabilities in the beam fluence we cannot account for. We do find, however, areas in the low-fluence grid where sites show few NV centers (typically one or none), hence suggesting the conversion efficiency is not far from typical values (of order 50% for bulk NVs[33]).

To experimentally determine the NV depth, we monitor the NV fluorescence as a function of the objective's position $z$ along the optical axis; a single mode 600–750 nm fiber (Thorlabs) is used in this measurement to attain maximum axial resolution. Extended Data Fig. 1f shows the result for NV$_A$ and NV$_B$ (respectively, the 'source' and target NVs in Fig. 1c of the main text): We identify the diamond surface from the initial fluorescence onset (black dashed line in Extended Data Fig. 1f) stemming from reflected laser light (leaking into the detector despite the filters) as well as from surface-defect-induced fluorescence. From the position of the second fluorescence maximum, we determine a depth of ~10 μm and ~9 μm for NV$_A$ and NV$_B$, respectively, slightly greater than — but still consistent with — the value obtained

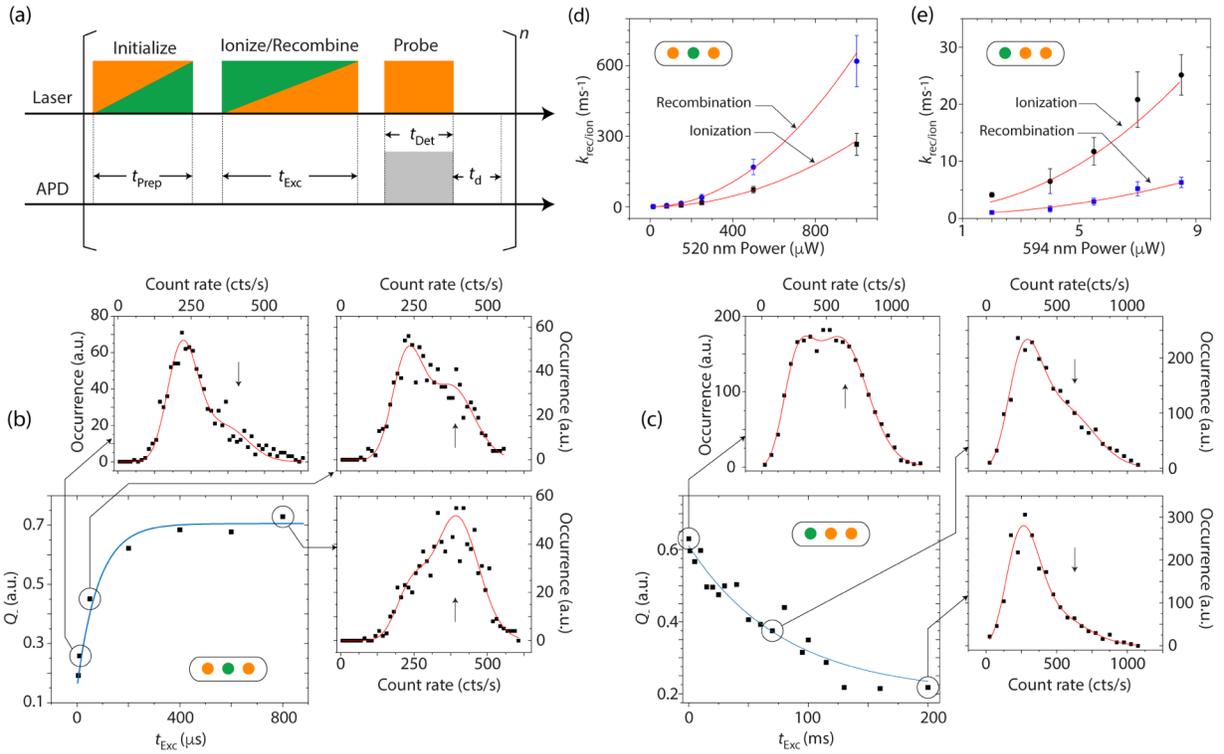

**Extended Data Fig. 2 | Charge control of individual NVs.** (a) Experimental protocol. Green (orange) blocks indicate laser excitation at 520 nm (594 nm). Left (right) triangles apply to the NV$^-$ recombination (ionization) measurements. APD denotes the avalanche photo-detector. (b) Relative NV$^-$ population as a function of the excitation time $t_{Exc}$ for 520 nm, 35 μW laser light; the solid trace is an exponential fit and the color code in the lower right corner indicates the laser color sequence. Surrounding plots are example histograms for different values of $t_{Exc}$, with solid red traces indicating fits corresponding to two Poissonian distributions with differing average count rates. Black arrows indicate contributions due to NV$^-$; both ionization and recombination rates at the excitation wavelength and power can be extracted from these fits (see text for details). During initialization, the orange laser power and duration are 7 μW and 200 ms, respectively; the readout pulse at 594 nm has a power of 7 μW and a duration of 80 ms. (c) Relative NV$^-$ population as a function of the excitation time $t_{Exc}$ produced by a 594 nm, 4 μW laser. As in (b), NV$^-$ populations can be extracted from analysis of the recorded histograms. In this protocol, the 520 nm initialization (594 nm readout) pulse has 200 μW (4 μW) power and a duration of 500 μs (20 ms). (d) NV recombination and ionization rates $k_{rec}$, $k_{ion}$ (blue and black dots, respectively) as a function of the 520 nm laser power as extracted from experiments similar to those in (b). Solid red traces indicate parabolic fits. (e) Same as in (d) but for 594 nm. All experiments are carried out in NV$_A$ ('source' NV in Fig. 2 of the main text). Similar results are obtained for NV$_B$, not shown here for brevity.

from SRIM.

An extensive sample characterization shows that NV sites resulting from the highest beam fluences — which, unfortunately, included all ring-like geometries — feature a charge state rather insensitive to optical excitation. In particular, ionization into a neutral state proved difficult or short lived, which suggests electron tunneling from nearby traps, as observed previously[12]. On the contrary, we find that lower NV density sites exhibit metastable (i.e., infinitely-lived) charge states that can be controlled optically. The experiments reported in this article are limited to this second class of sites.

**Charge state control of individual NVs**

To properly interpret our carrier transport experiments, we first proceed to characterize the individual charge state dynamics of the source and target NVs (Fig. 1c in the main text). Extended Data Fig. 2a shows our experimental protocol: We use consecutive green (520 nm) and orange (594 nm) laser pulses to initialize and subsequently transform the NV charge state; we then conduct a charge-state preserving NV readout via a low power orange pulse. Since strong green (orange) illumination preferentially prepares the NV in the negative (neutral) charge state, the ensuing orange (green) excitation pulse predominantly leads to NV$^-$ ionization (NV$^0$ recombination). Note, however, that although one process or the other prevails depending on the laser wavelength[7], both are invariably present, meaning that an integrated analysis must be put in place to properly extract the underlying charge conversion rates (see below). On the other hand, orange excitation is known to have a comparatively lower impact on the NV charge state (a property we exploited in the main text to observe charge state jumps in real time, Fig. 1c), thus justifying the use of weak orange excitation for charge state readout.

To identify the fractional NV populations in either charge state, we plot our results in the form of histograms, each representing the probability distribution associated to fluorescence readouts with a given photon count. Extended Data Figs. 2b and 2c show some examples corresponding to alternative variants of the measurement protocol (see color code inserts in the main plots): In all cases, we identify a bimodal dispersion whose low-count-rate (high-count-rate) peak reveals the fractional NV$^0$ (NV$^-$) population at a given time during the charge-conversion segment $t_{Exc}$ of the protocol. We extract either relative population from a fit containing two Poissonian distributions with distinct average photon count rates. Extended Data Figs. 2b and 2c display the evolution of the NV$^-$ population as a function of $t_{Exc}$ under laser excitation at 520 nm and 594 nm, respectively. The former indicates the dynamics are dominated by NV$^0$ recombination while the latter mainly signals NV$^-$ ionization; in both instances, however, the charge conversion is partial, hence confirming the simultaneous presence of a counter mechanism.

Formally, we describe local NV charge conversion via a set of master equations, namely:

$$\begin{aligned}\frac{dQ_-}{dt} &= -k_{ion}Q_- + k_{rec}Q_0 \\ \frac{dQ_0}{dt} &= k_{ion}Q_- - k_{rec}Q_0\end{aligned} \quad (1)$$

Here, $Q_-$ and $Q_0$ correspondingly denote the normalized NV$^-$ and NV$^0$ populations; $k_{ion}$ and $k_{rec}$ respectively stand for the ionization and recombination rates under photoexcitation. With the simplifying assumption of full NV$^-$ initialization under green excitation, the detailed balance set in Eq. 1 has the following solutions:

$$\begin{aligned}Q_-(t) &= \frac{k_{rec}}{k_{ion}+k_{rec}} + \frac{k_{ion}}{k_{ion}+k_{rec}}e^{-(k_{rec}+k_{ion})t} \\ Q_0(t) &= \frac{k_{ion}}{k_{ion}+k_{rec}} - \frac{k_{ion}}{k_{ion}+k_{rec}}e^{-(k_{rec}+k_{ion})t}\end{aligned} \quad (2)$$

While the time constant alone is insufficient to separate contributions from NV recombination and ionization, the characteristic rate of both processes impacts the amplitudes of the transient and time-independent components of the response, thus allowing us to extract either rate from a fit. Extended Data Figs. 2d and 2e show the results derived from this method for 520 and 594 nm excitation of variable power. All rates display a quadratic dependence within the probed range, consistent with the two-photon nature of either process[7]. Interestingly, the power dependence of the $NV^-$ →$NV^0$ conversion rate of the target due to carrier capture (insert in Fig. 2d of the main text) shows a quadratic power dependence as well, which supports the notion that captured carriers are indeed being photo-ionized out of a source NV under 520 nm illumination. These measurements give us a precise knowledge of the charge carrier generation rate out of the source NV, an instrumental piece of information for developing and quantifying a model of charge capture by a single defect (see below).

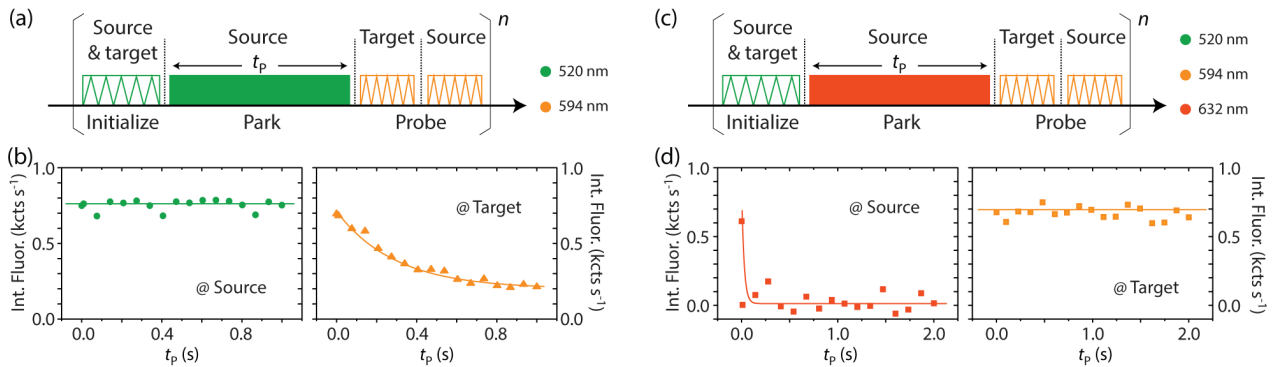

**Extended Data Fig. 3 | Source-target carrier transport at different excitation wavelengths.** (a) Working with the NV pair presented in Fig. 2 of the main text, we modify the experimental protocol to include a 594 nm fluorescence scan over a 1.5×1.5 μm$^2$ area around the source NV, which we use to determine its integrated fluorescence. (b) Time-averaged, integrated fluorescence of the source and target NVs (left and right panels, respectively); while the former gradually bleaches, the latter remains bright. The 520 nm laser intensity during the park is 1.0 mW; all other conditions as in Fig. 2. (c) Same as in (a) but using a 1.5 mW, 632 nm park. (d) Same as in (b) but for the protocol in (c). The source NV transitions immediately to a dark state while the target remains bright. Solid lines in (b) and (d) are guides to the eye.

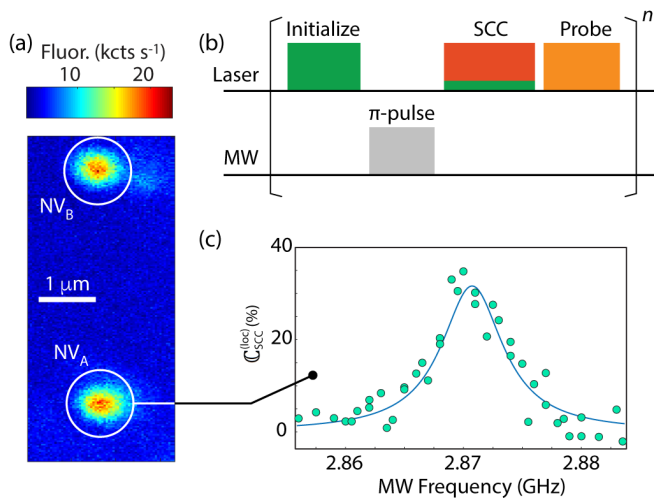

**Extended Data Fig. 4 | Local spin-to-charge conversion.** (a) Fluorescence image of target and source NVs. (b) Spin-to-charge conversion (SCC) protocol. Green, red, and orange blocks indicate laser excitation at 520 nm, 632 nm, and 594 nm, respectively. During SCC, red and green excitation take place simultaneously. The grey block indicates MW excitation at variable frequency; the MW pulse length and amplitude are chosen to induce a population inversion when on-resonance with the NV ground state zero-field splitting (~2.87 GHz). (c) Fluorescence change from $NV_A$ (source) as a function of the MW frequency upon application of the SCC protocol in (b). Throughout these experiments, the power and duration of the initialization (probe) pulse are 200 μW and 500 μs (7 μW and 15 ms), respectively. The green (red) laser power during SCC is 200 μW (16 mW); the SCC pulse duration amounts to 300 ns.

Another important piece of information follows from the experiments in Extended Data Fig. 3, where we simultaneously monitor the fluorescence response from the NVs in Fig. 2 after laser parks at the source using two different wavelengths. Consistent with the model of cyclic charge generation and transport, we find that 532 nm optical excitation dims the target NV but leaves the source brightness unchanged. Conversely, optical excitation at 632 nm immediately bleaches the source NV but has no impact on the target. An immediate conclusion is that if one was to assume the existence of unaccounted carrier sources, their response to optical excitation of variable wavelength must be similar to that of an NV.

**Local spin-to-charge conversion**

From the spin-dependent transport experiments in Fig. 3 of the main text we conclude that a fraction of the carriers that contribute to converting the charge state of the target defect originate from the source NV. In order to quantitatively determine how large this fraction is, we conduct local spin-to-charge conversion measurements at the source NV site following the protocol shown in Extended Data Fig. 4b. Our protocol relies on simultaneous green and red excitation to excite and subsequently trigger ionization of $NV^-$ as described previously[35]. We adjust the relative powers empirically for optimal SCC contrast as the process involves a subtle interplay between NV ionization and recombination. Importantly, the spin-to-charge conversion (SCC) pulse is identical (i.e., same duration and laser powers) to that used in Fig. 3 of the main text. Extended Data Fig. 4c shows the source NV response as a function of the MW frequency: Consistent with prior SCC work in individual NVs[21], we attain a conversion contrast of ~30%.

To correlate this result with the observations in the main text (Fig. 3d), we first note that because an NV center — be it in the neutral or negative state — undergoes a charge conversion cycle during optical excitation, the emission of a free electron upon $NV^-$ ionization must be necessarily compensated with the injection of a hole during $NV^0$ recombination. Correspondingly, the fluorescence change in the target reflects on the contrast attained locally at the source upon completion of a single SCC cycle (even if the overall signal-to-noise ratio decreases due to the probabilistic nature of the carrier exchange between NVs[53]). In particular, comparison with the contrast obtained in Fig. 3 of the main text (reaching up to ~20-25%) implies that a fraction greater than 75% of the charge carriers that are trapped by the target NV are photo-ionized out of the source NV. We stress this is a lower bound; future work — for example, in the form of experiments at variable excitation wavelengths — will allow us to set a more precise correspondence and, if present, shed light on the nature of any parasitic source.

**Carrier transport between few-NV islands**

The observations in Figs. 2 and 3 of the main text are not exclusive to individual NVs. An example is shown in Extended Data Fig. 5, where we investigate transport between sites comprising two and three NVs (lower and upper circles in Extended Data Fig. 5a, respectively). As observed for single NVs (and ensembles[9]), prolonged green laser parking at the source site (protocol in Fig. 2b of the main text) leads to gradual bleaching of the NVs at the target site (Extended Data Fig. 5b). By the same token, application of the spin-to-charge conversion in Fig. 3a of the main text, allows us to encode the (common) spin state of the source NVs into the (common) charge state of the target set; further, by varying the MW frequency, we reconstruct a charge encoded magnetic resonance spectrum of the source, analogous to that observed in Fig. 3d of the main text.

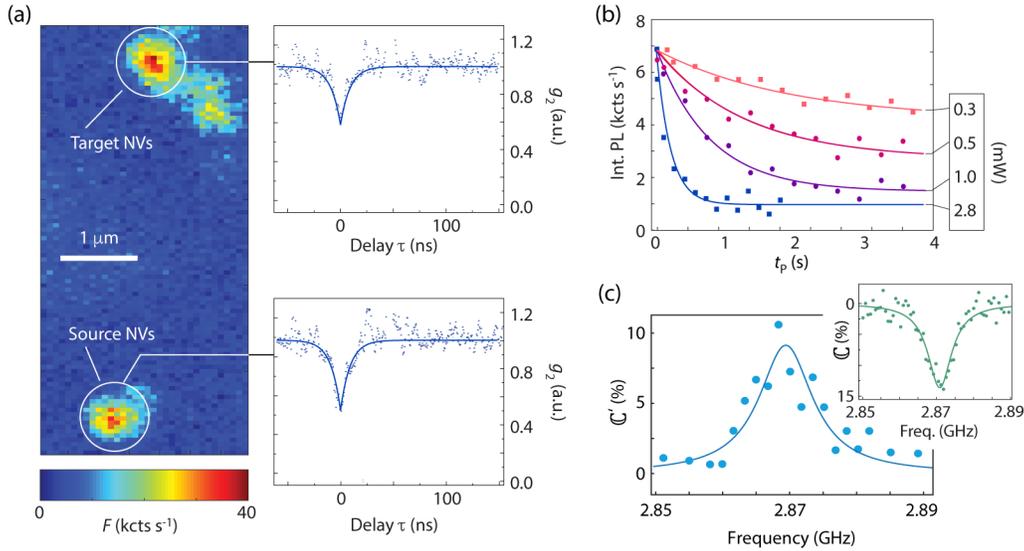

**Extended Data Fig. 5 | Transport between few-NV islands.** (a) Fluorescence image of two proximal ion implantation spots. Time-correlated single-photon measurements at the circled sites indicate the presence of 2 or 3 NVs at each location. (b) NV⁻ photoluminescence at the target site upon application of the protocol in Fig. 2b of the main text for variable park times $t_P$ and different laser powers. (c) NV⁻ fluorescence from the target site as a function of the MW frequency upon spin-to-charge conversion at the source site (protocol in Fig. 3a of the main text). As a reference, the insert the displays the optically detected magnetic resonance spectrum of the source NVs. Conditions are similar to those in Figs. 2 and 3 of the main text.

**The impact of externally applied electric fields**

While the experiments in the main text rely on carrier self-diffusion from the point of injection, the application of external electric fields should allow us to gain some degree of control over the transport process. To explore this possibility, we pattern metal electrodes on the sample substrate in the form of metal pads separated by a 500 μm gap; the system is oriented so that the resulting electric field **E** be parallel to the line connecting the source and target NVs (Extended Data Fig. 6a). To gauge the impact of the field on the inter-NV carrier transport, we implement a protocol analogous to that in Fig. 2 of the main text but adapted to include a voltage pulse of variable amplitude synchronous with the green laser park (Extended Data Fig. 6b). Extended Data Fig. 6c displays the target NV response as a function of the laser park duration for different voltages. Greater fields gradually lead to slower fluorescence bleaching rates hence indicating progressive blocking of carrier transport from the source; in particular, we observe no change in the target NV charge state for voltages above ~60 V (corresponding to electric fields $E \approx 120$ mV μm⁻¹).

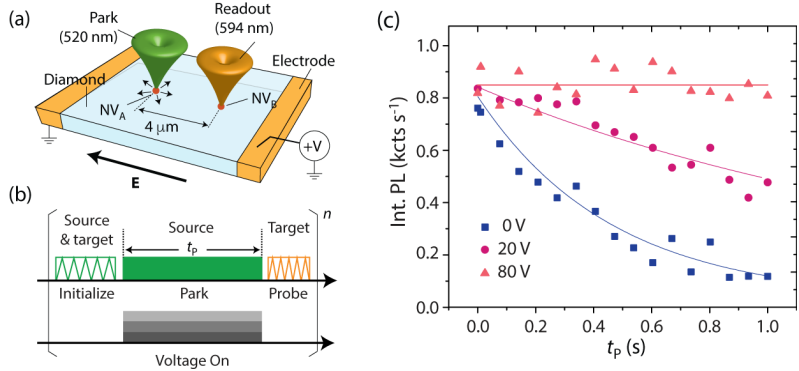

**Extended Data Fig. 6 | Inter-NV transport in the presence of externally applied electric fields.** (a) Schematics of the experimental setup. We pattern metal pads on the sample substrate to produce an electric field **E** parallel to the axis connecting the source and target NVs (respectively, NV$_A$ and NV$_B$). The gap between the electrodes is 500 μm. (b) Pulse protocol. After NV initialization (green scan at the source and target NVs), we apply a voltage of variable (but fixed) amplitude synchronically with the green laser park at the source NV; we readout the charge state of the target NV via an orange scan. Positive voltages create an electric field pointing from NV$_B$ to NV$_A$. (c) Integrated fluorescence of the target NV as a function of the green laser park time $t_P$ for different voltages. The green laser power during the park is 1 mW; all other conditions as in Fig. 2 of the main text.

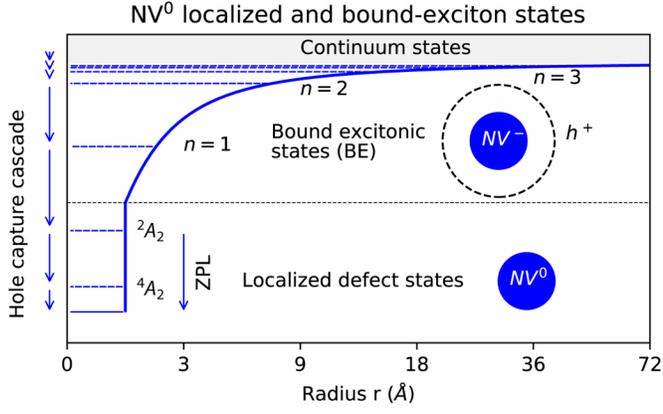

**Extended Data Fig. 7 | Cascade capture of a hole by a negatively charged NV.** We consider the formation of a bound exciton state comprising a hole orbiting an NV⁻ core. This system can be described by a hydrogenic series of states characterized by a quantum number $n$. Assuming the hole is in thermal equilibrium, bound exciton formation is possible for orbits $n_t$ such that $(E_I - E_n) \sim \kappa_B T$; subsequent capture involves the cascade emission of phonons. The corresponding trapping radius (also known as the 'Onsager' radius[58,54]) is given by $r_t = e^2/(4\pi\varepsilon\kappa_B T) \approx$ 10 nm in room temperature diamond. The ensuing capture cross section would then amount to $\sigma_{h,On} \sim 3\times 10^{-4}$ μm², not far off from the experimental value.

Throughout these measurements, we also observe strong hysteresis in the system response that we attribute to space charge fields stemming from the accumulation of trapped carriers at the interface with the electrodes. Recent experiments show these fields can be quite strong — exceeding several V μm⁻¹ — though their exact amplitude and spatial geometry depend sensitively on the relative timing and wavelength of the laser (or lasers) used for photo-ionization[38]. In particular, we have already shown these fields can be exploited for carrier guiding[38], though additional work will be needed to gain a fuller understanding.

**Extracting the NV⁻ hole capture cross section**

In the presence of sustained carrier injection from the source, the charge state dynamics of the target NV can be cast as

$$\frac{dQ_-^{(tar)}}{dt} = \rho_h \xi_h Q_-^{(tar)}, \quad (3)$$

where $\rho_h$ is the hole density at the target site, $\xi_h$ is the NV⁻ hole capture coefficient, and we have neglected trapping of electrons by NV⁰. Assuming radial hole injection from the source with rate $k_h$, charge conservation arguments allow us to express the hole density at the target site as

$$\rho_h = \frac{k_h}{4\pi d^2 v_h}, \quad (4)$$

where $d$ denotes the source–target distance, and $v_h$ is the hole velocity. Since $\xi_h = \sigma_h v_h$ — i.e., the product between the NV⁻ hole capture cross section $\sigma_h$ and the hole velocity $v_h$ — we write the solution to Eq. 3 as $Q_-^{(tar)}(t) = Q_-^{(tar)}(0)e^{-t/\tau}$ where $Q_-^{(tar)}(0) \approx 0.75$ and

$$\tau = \frac{4\pi d^2}{\sigma_h}\left(\frac{1}{k_{ion}} + \frac{1}{k_{rec}}\right), \quad (5)$$

represents the target NV⁻ ionization time constant in the presence of hole transport. In Eq. (5), we assume the hole injection rate is given by $k_h \sim \left(\frac{1}{k_{ion}} + \frac{1}{k_{rec}}\right)^{-1}$ (roughly, the inverse of the time needed for a full cycle of NV ionization and recombination). Since $k_{ion}$ and $k_{rec}$ under 532 nm illumination are known quadratic functions of the laser power (see Extended Data Fig. 2d), we use Eq. (5) to simultaneously fit the target NV⁻ response observed in Figs. 2d and 4c of the main text. We obtain $\sigma_h \sim 3\times 10^{-3}$ μm², orders of magnitude greater than typical values reported for ensemble measurements in comparable systems[41].

An arguably more intuitive interpretation of our results emerges from considering the number of emitted holes necessary to induce bleaching of the target fluorescence. For example, from Extended Data Fig.2d, we roughly estimate the hole injection rate is $\sim 10^6$ s⁻¹ at 2 mW of 520 nm laser (being the slower of the two, the ionization rate determines the overall rate of an optical ionization and recombination cycle). On the other hand, the charge state conversion time by the target NV under 2 mW, 520 nm laser illumination of the source is $\sim 0.1$s (Fig. 2d). Therefore, we conclude that one hole is captured by the target NV out of $10^5$ holes emitted by the source NV during the charge transfer protocol. For an inter-defect distance of $\sim 4$ μm — the case in Fig. 2 — this $\sim 10$ ppm fraction roughly amounts to the capture cross section calculated above.

Lastly, we mention that the presence of charge traps other than the target NV — ignored in the calculations above — can only extend the time required to observe hole capture by the target, implying that the calculated cross section must be seen as a lower bound. On the other hand, we have observed that optical excitation at sites other than those implanted has no effect on the fluorescence of the target NV, hence suggesting the overall concentration of charge traps is low (at least, in terms of the impact they have in the present experiments). Additional work will be needed, however, to attain a quantitative estimate.

**Model of hole capture by NV⁻**

Following Ref [40], we hypothesize the capture of a hole by NV⁻ in diamond starts with the transient formation of a bound exciton state in which the hole occupies an outer hydrogenic orbit centered at the negatively charged defect (see Extended Data Fig. 7). The existence of such bound exciton states has already been exposed in SiV⁻ centers in diamond via optical spectroscopy[39] and, as our ab-initio calculations show, they should also be present in the case at hand (see below).

In the simplest representation, the Hamiltonian for an NV⁻-bound exciton reads

$$H = -\frac{\hbar^2}{2m^*}\nabla^2 - \frac{e^2}{4\pi\varepsilon r}, \quad (6)$$

where $m^*$ denotes the effective mass of the hole, and $\varepsilon$ is the dielectric constant of diamond. Diagonalization of this Hamiltonian leads to the hydrogenic series of energy states given by

$$E_n = E_I - \frac{\varepsilon_0^2}{\varepsilon^2}\frac{m^*}{m_e}\frac{E_{Ry}}{n^2}, \quad (7)$$

where $E_I$ represents the ionization energy, $E_{Ry} = 13.6$ eV is the Rydberg energy, and $m_e$ is the free electron mass. Qualitatively, one would anticipate hole capture when $E_I - E_n \sim \kappa_B T$, which translates into the Onsager trapping radius

$$r_t = \frac{e^2}{(4\pi\varepsilon\kappa_B T)}. \quad (8)$$

In the above formula, $\kappa_B$ denotes Boltzmann's constant, $T$ is the absolute temperature, and we made use of the expression $r_n = \frac{\varepsilon}{\varepsilon_0}\frac{m_e}{m^*}n^2 a_0$ for the effective radius of the $n$-th orbital, with $a_0 = 0.053$ nm denoting the Bohr radius. Notice that the orbital radii grow with the dielectric constant and depend inversely on the effective carrier mass, which renders $r_t$ ultimately insensitive to $m^*$.

Using DFT at the HSE06 level as outlined below, we find values for $E_I = 2.71$ eV, and $\varepsilon = 5.83\,\varepsilon_0$. For the effective masses of the hole that can be present either in the light-hole (*lh*), heavy-hole (*hh*), or split-off configurations (*so*), we use values reported previously[55,56], namely, $m_{hh}^{[111]} = 0.788\,m_e$, $m_{hh}^{[100]} = 0.366\,m_e$, $m_{hh}^{[110]} = 1.783\,m_e$, $m_{lh}^{[111]} = 0.788\,m_e$, $m_{lh}^{[100]} = 0.366\,m_e$, $m_{lh}^{[110]} = 0.366\,m_e$, $m_{so}^{[111]} = 0.198\,m_e$, $m_{so}^{[100]} = 0.466\,m_e$, $m_{so}^{[110]} = 0.232\,m_e$. In Table 1, we show the corresponding values for the excitation energy and the radii exemplified along the [110] direction. For the room-temperature Onsager radius, we obtain $r_t \sim 10$ nm, and, correspondingly, we estimate the hole capture cross section as $\sigma_{h,On} \sim \pi r_t^2 \sim 3\times 10^{-4}$ μm$^2$, within an order of magnitude of the observed value ($\sigma_h = (3\pm 1)\times 10^{-3}$ μm$^2$).

| $n$ | $E_n^{lh}$ (eV) | $r_n^{lh}$ (nm) | $E_n^{hh}$ (eV) | $r_n^{hh}$ (nm) | $E_n^{so}$ (eV) | $r_n^{so}$ (nm) |
|---|---|---|---|---|---|---|
| 1 | 2.564 | 0.843 | 1.997 | 0.173 | 2.617 | 1.330 |
| 2 | 2.673 | 3.372 | 2.532 | 0.692 | 2.690 | 5.319 |
| 3 | 2.694 | 7.587 | 2.631 | 1.557 | 2.699 | 11.969 |
| 4 | 2.700 | 13.487 | 2.665 | 2.769 | 2.704 | 21.277 |

**Extended Data Table 1 | Excitation energies and radii.** We consider the lowest four bound exciton states along the [110] direction based on Eq. (7).

A more rigorous approach to determining the hole capture cross section is the cascade trapping framework developed by M. Lax[40,57]. Building on a semi-classical picture, this model describes a process where the energy accrued by the carrier as it accelerates towards the charged defect at the core is released via the emission of consecutive phonons[58]. The Lax model identifies two different regimes: In the *Thompson limit,* the underlying assumption is that the carrier mean free path is long compared with the characteristic trapping distance $2r_t$, so that one or fewer collisions of the hole occur close to the negative charge center prior to the cascade process. By contrast, in the *Langevin limit* the free carrier dynamics is diffusive, i.e., many collisions occur close to the charge center.

First assuming that the Thompson regime applies, we consider two alternative cases where the cascade process is dominated either by acoustic or by optical phonons; the latter should be dominant at higher temperatures while the former should prevail under cryogenic conditions. Following Ref. [40] and under the assumption of a hole mobility[59] of 1200 cm$^2$V$^{-1}$s$^{-1}$, a speed of sound in diamond[60] of 16000 m s$^{-1}$, and the effective hole masses introduced above, we find that the acoustic-phonon-driven capture cross section has a value of $\sigma_{h,Th}^{ac} = 1.7\times 10^{-6}$ μm$^2$. Analogously, we obtain $\sigma_{h,Th}^{op} = 1.2\times 10^{-5}$ μm$^2$ for the case of scattering by optical phonons, which should be valid at room temperature. Here we employed $\hbar\omega_{op} = 165$ meV as the corresponding optical phonon energy[61]. While both results are considerably lower than the experimental value, we note that a formalism where semi-classical trajectories are replaced by a distribution function would likely yield greater cross sections[57].

Finally, we can also use Lax' framework to derive a capture cross section valid for the *Langevin regime*: We obtain $\sigma_{h,La} = 4.35\times 10^{-3}$ μm$^2$, consistent with the experimental result. While the close agreement might be partly serendipitous, this regime seems justified given the known phonon scattering rates for carriers in room-temperature diamond[42,43], of order $\sim 7.5\times 10^{12}$ s$^{-1}$, corresponding to a mean free path $l \sim 15$ nm comparable to the trapping range $2r_t$.

To characterize the NV-bound excitons in diamond from first principles, we use spin-polarized density functional theory (DFT) as implemented in the VASP 6.1.0 code[62] with the PBE[63] and the HSE06[64] exchange-correlation functionals. For the PBE functional, we consider seven different diamond supercells comprising 4096, 2744, 1728, 1000, 512, 216, and 64 atoms. These supercells have been created by converging the cubic diamond unit cell containing 8 atoms with 12×12×12 *k*-points leading to a lattice constant of 0.356 nm. Using these values, we construct the supercells including the NV center and optimize the structures until forces are less than $10^{-3}$ eV nm$^{-1}$ with plane wave energy cutoff of 420 eV. Due to the high computational costs of the HSE06 functional, we optimized supercells including 1000, 512, 216, and 64 atoms on this level of theory. Extended Data Fig. 8a shows the calculated charge transition level of the (-1/0) charge state by using an extrapolation of different supercell sizes. This energy corresponds to the ionization energy of the NV- center, as well as to the $n \to \infty$ limit in the exciton model introduced above. For PBE and HSE06, we compare two different situations, with and without a charge correction scheme[65]. Upon interpolation to infinite carbon atom count, we find that the corrected and uncorrected schemes lead to the same (-1/0) charge transition level, namely, 2.05 eV and 2.71 eV for the PBE and HSE06 functionals, respectively. The HSE06 value is in agreement with the experimental value of 2.6 eV[7].

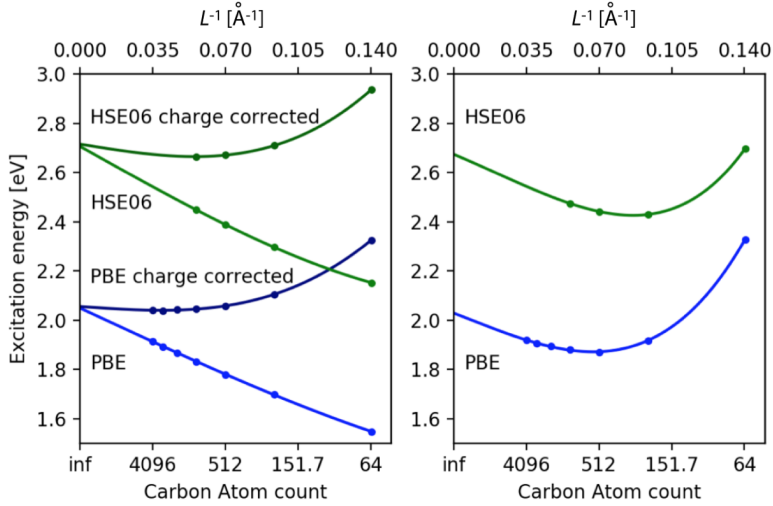

**Extended Data Fig. 8 | DFT modelling.** (Left) PBE- and HSE06-calculated exchange-correlation functional for the (0/-1) charge transition level (blue and green dots, respectively), which corresponds to the ionization energy of the NV- center. For each functional, we show two values, the uncorrected (light) and the charge corrected[65] (dark) excitation energies. Both lead to the same value for the extrapolation at infinite carbon atoms in the supercell. (Right) $n = 1$ bound exciton energy using the HSE06 (green) and the PBE (blue) exchange-correlation functional. The solid lines in (a) stem from fits using the formula $f(x) = a + (bx + cx^3)\exp(-dx)$, where $a, b, c$, and $d$ are fitting parameters, and $x \equiv L^{-1}$ is the inverse supercell length. All values are listed on Table 2.

While the long-range of the Coulomb potential makes present computing power insufficient for a full ab-initio characterization of the bound exciton dynamics, it is nonetheless possible to assess the feasibility of the proposed NV model by investigating the $n = 1$ state, where confinement is largest (Extended Data Fig. 8b). To this end, we follow the DeltaSCF method outlined in Ref. [39]. The procedure consists of two steps: We first perform a ground-state total energy calculation for $NV^0$ to subsequently obtain an excited state energy by promoting an electron from the delocalized valence bands to a state localized at the NV; the electronic configuration of the localized states resembles the configuration of the NV- center. This procedure leads to a $n = 1$ bound exciton energy of 2.03 eV PBE and 2.67 eV HSE06, both lower than their respective ionization energies shown in Extended Data Fig 7a.

Despite the present limitations, extensions are possible, for example, in the form of effective models amenable to larger system sizes[66], and more accurate many-body methods, such as Green's functions and Bethe-Salpeter-like calculations[67].

**Data availability**

The data that support the findings of this study are available from the corresponding author upon reasonable request.

| NV (0, -1) transition energy | | | | |
|---|---|---|---|---|
| $f(x) = a + (bx + cx^3)$ | $a$ (eV) | $b$ (eV Å) | $c$ (eV Å$^2$) | |
| PBE | 2.05 | -3.93 | 18.03 | |
| PBE (charge corrected) | 2.06 | -0.59 | 127.67 | |
| HS06 | 2.71 | -4.74 | 40.00 | |
| HS06 (charge corrected) | 2.72 | -1.37 | 149.67 | |
| $n=1$ bound exciton energy | | | | |
| $f(x) = a + (bx + cx^3)\exp(-dx)$ | $a$ (eV) | $b$ (eV Å) | $c$ (eV Å$^2$) | $d$ (eV Å) |
| PBE | 2.03 | -3.30 | 257.44 | 1.34 |
| HS06 | 2.67 | -3.57 | 186.27 | 3.23 |

**Extended Data Table 2:** Fitting parameters for the DFT data shown in Extended Data Fig. 8. In these equations $x \equiv L^{-1}$ is the inverse supercell length.

**Acknowledgments**. The authors acknowledge useful discussions with Y.H. Chen. A.L., J.H., and C.A.M acknowledge support from the National Science Foundation through grant NSF-1914945, and from Research Corporation for Science Advancement through a FRED Award; they also acknowledge access to the facilities and research infrastructure of the NSF CREST IDEALS, grant number NSF-HRD-1547830. M.W.D. acknowledges support from the Australian Research Council COE170100169. Ion implantation work to generate the NV and SiV centers was performed, in part, at the Center for Integrated Nanotechnologies, an Office of Science User Facility operated for the U.S. Department of Energy (DOE) Office of Science. Sandia National Laboratories is a multimission laboratory managed and operated by National Technology & Engineering Solutions of Sandia, LLC, a wholly owned subsidiary of Honeywell International Inc., for the U.S. Department of Energy's National Nuclear Security Administration under contract DE-NA0003525.This paper describes objective technical results and analysis. Any subjective views or opinions that might be expressed in the paper do not necessarily represent the views of the DOE or the United States Government. The Flatiron Institute is a division of the Simons Foundation.

**Author contributions**. A.L., H.J., and C.A.M. conceived the experiments. G.V. and E.B. ion-implanted the sample. A.L. conducted the experiments and analyzed the data with assistance from H.J., D.D., and C.A.M.; J.F. led the theoretical and numerical modelling with assistance from M.W.D.; C.A.M., A.L., and J.F. wrote the manuscript with input from all authors. C.A.M. supervised the project.

**Competing interests**. The authors declare no competing interests.

**Additional information**

**Correspondence and requests for materials** should be addressed to C.A.M.

**Reprints and permissions** information is available at http://www.nature.com/reprints.